\documentclass[usenatbib]{mn2e}
\usepackage{epsfig}
\usepackage{colordvi}
\usepackage{graphicx}
\usepackage{amssymb,textcomp}
\usepackage{color}

\newcommand{\apjs}{ApJS}
\newcommand{\apj}{ApJ}
\newcommand{\apjl}{ApJ}

\newcommand{\aap}{A\&A}

\newcommand{\mnras}{MNRAS}
\newcommand{\nat}{Nature}
\newcommand{\nar}{New Astron. Rev.}
\newcommand{\araa}{ARA\&A}
          
\newcommand{\aplett}{Astrophys. Lett.}          
          
\newcommand{\rmd}{{\rm d}}       
\newcommand{\msun}{\mbox{M}_{\sun}}    
\newcommand{\rsun}{\mbox{R}_{\sun}}    
 
\newcommand {\beq}{\begin {eqnarray}}
\newcommand {\eeq}{\end {eqnarray}}

\newcommand{\ergs}{erg~s$^{-1}$}   

\newcommand{\source}{{XTE~J1550--564}}   

\title[Colours of  black holes]
{Colours of  black holes: 
infrared flares from the hot accretion disc in \source}
 \author[J. Poutanen, A. Veledina and M.G. Revnivtsev]
    {Juri~Poutanen$^{1,2}$\thanks{E-mail: juri.poutanen@utu.fi} 
  Alexandra~Veledina$^{2,1}$ 
  and Mikhail G. Revnivtsev$^{3}$
  \\
$^1$Tuorla Observatory, Department of Physics and Astronomy, University of Turku, V\"ais\"al\"antie 20, FI-21500 Piikki\"o, Finland\\
$^2$Astronomy Division, Department of Physics, PO Box 3000, FI-90014 University of Oulu, Finland \\
$^3$Space Research Institute, Russian Academy of Sciences, Profsoyuznaya 84/32, 117997 Moscow, Russia\\
}

\begin{document}

\pagerange{\pageref{firstpage}--\pageref{lastpage}}
\pubyear{2014}
\date{Accepted 2014 September 22.  Received 2014 September 3; in original form 2014 June 29}

\maketitle
\label{firstpage} 

\begin{abstract}
\noindent
Outbursts of the black hole (BH) X-ray binaries are dramatic events occurring in our Galaxy approximately once a year.
They are detected by the X-ray telescopes and often monitored at longer wavelengths.
We analyse the X-ray and optical/infrared (OIR) light-curves of the BH binary \source\ during the 2000 outburst. 
By using the observed extreme colours as well as the characteristic decay time-scales of the OIR and X-ray light curves,
we put strong constraints on the extinction towards the source. 
We accurately separate the contributions to the OIR flux of the irradiated accretion disc  and a non-thermal component.
We show that the OIR non-thermal component appears during the X-ray state transitions both during 
the rising and the decaying part of the outburst at nearly the same X-ray hardness but at luminosities differing 
by a factor of 3. 
The line marking  the quenching/recovery of the non-thermal component at the X-ray hardness--flux diagram  
seems to coincide with the `jet line' that marks the presence of the compact radio jet.  
The inferred spectral shape and the evolution of the non-thermal component  
during the outburst, however, are not consistent with the jet origin, but  are naturally explained in terms of the hybrid hot flow scenario,
where  non-thermal electrons emit synchrotron radiation  in the OIR band. 
This implies a close, possibly causal connection between the presence of the hot flow and the compact jet. 
We find that the non-thermal component is hardening during the hard state at the decaying stage of the outburst, which indicates 
that the acceleration efficiency is a steep function of radius at low accretion rate. 
\end{abstract}


\section{Introduction}

The optical/infrared (OIR) spectra of black hole (BH) low-mass X-ray binaries often show an excess above the standard \citep{SS73} accretion disc emission 
(e.g., \citealt{HMH00,HHC02}; \citealt*{GGH10}).
In some cases, the spectrum can be described by a power law of index close to zero (i.e. $F_\nu\propto \nu^0$).
There are three possible candidates that may account for this emission: 
the irradiated disc \citep{Cun76,GDP09}, hot accretion flow \citep{VPV13} and the jet \citep{HHC02,GMM07}.
Sometimes the OIR fluxes are higher than expected from any candidate alone \citep{CHM03,GDD10}, and the complex optical/X-ray cross-correlation functions \citep{Kanbach01,DGS08} support this anticipation, suggesting contribution of two components simultaneously \citep*{VPV11}.
The source of OIR emission cannot be determined by only using photometric data and 
some additional information about the OIR--X-ray connection, short time-scale variability properties and 
the long-term spectral variations is required (see review in \citealt{PV14}).
The latter is particularly important when trying to separate emission of different components.
 
\begin{figure*}
\centerline{ \epsfig{file=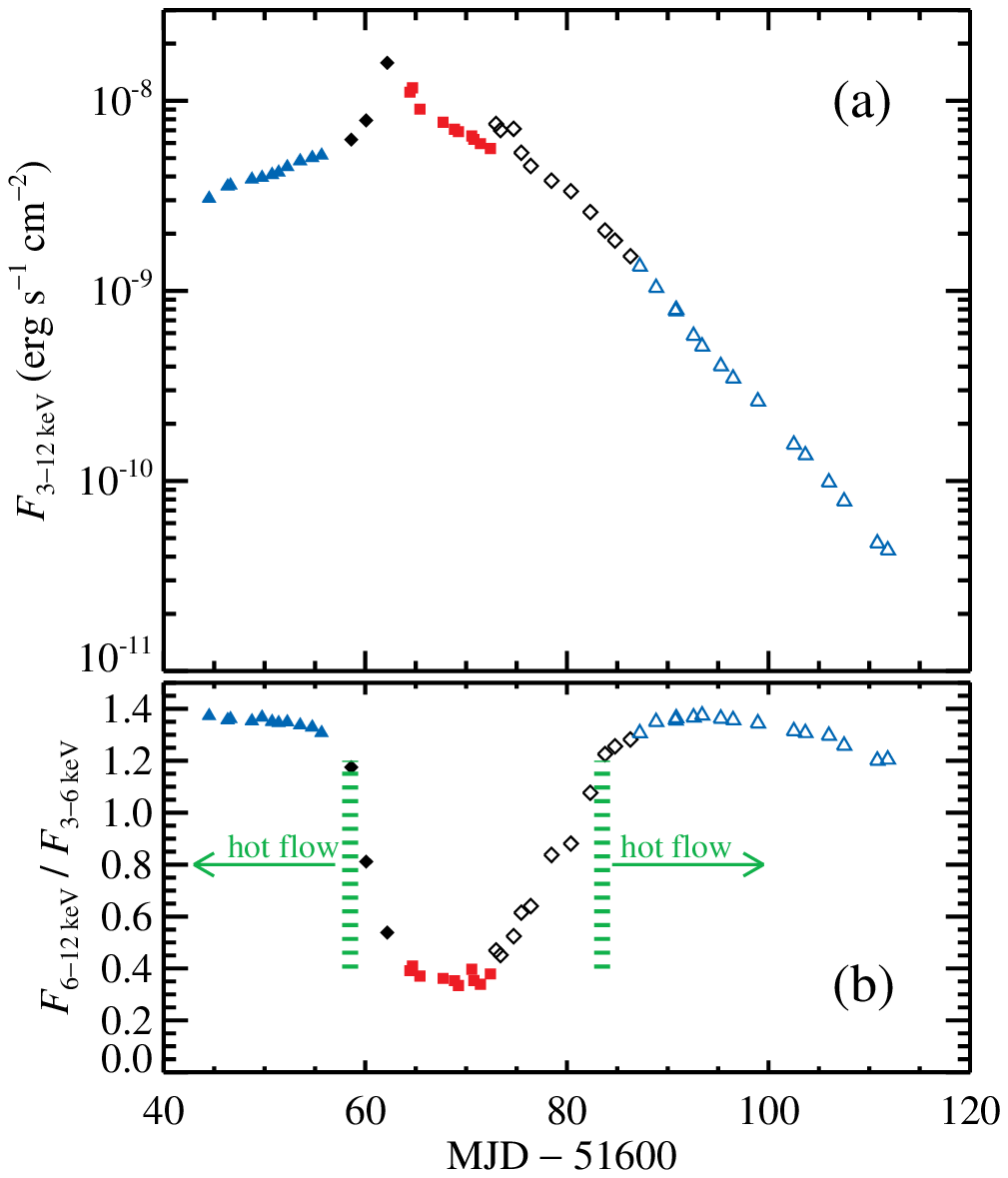, width=6.5cm} \hspace{0.5cm} 
 \epsfig{file=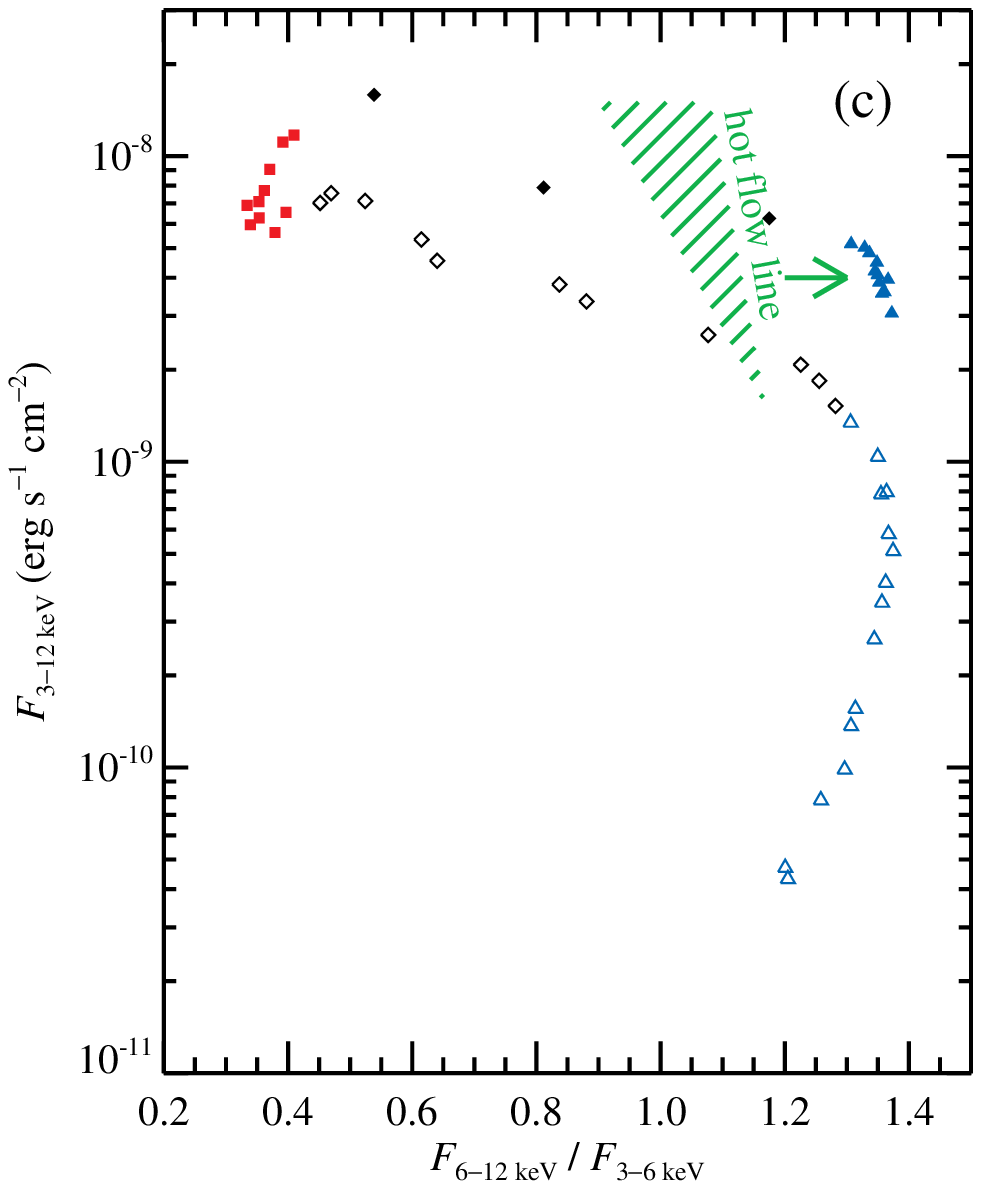, width=6.5cm}}
\caption{(a) Light curve of \source\ (flux in the 3--12 keV band) and (b) the evolution of the hardness ratio 
(i.e. ratio of fluxes in the energy bands 6--12  and 3--6 keV) during the 2000 outburst. 
(c) The hardness--flux diagram. Different symbols and colours highlight outburst stages defined  
from the hardness ratio (see text). 
The times and the positions of quenching/recovery of  the non-thermal OIR component (`hot flow line') are marked by green ribbons.
}
\label{fig_xrays_lc}
\end{figure*}

The entire transition of the BH transient \source\ from the hard to the soft state and back during its 2000 outburst was monitored 
by Yale 1 m telescope at CTIO in the $V$, $I$ and $H$ filters \citep{Jain01b}.
The light-curve structure cannot be simply explained by the fast rise-exponential decay pattern, expected in the case of standard or irradiated disc.
An additional component, manifesting itself through strong flares, is required. 
Recently, \citet{RMDM10,RMDF11} suggested that this component originates in a radio-emitting  jet;
however, multiple errors in these works make this interpretation doubtful. 
Here, we re-analyse the available OIR and X-ray data.
From the characteristic decay time-scales in the X-rays and 
in different OIR filters, we infer the typical  accretion disc temperature during the soft state.
This immediately translates to the constraints on the extinction towards the source, poorly known before. 
Using the OIR light curves, 
we accurately extract the non-disc non-thermal component and show evolution of its spectral shape during the flare.
We find that the additional component may originate in the hot accretion flow, if a small fraction of energy 
is injected in the form of non-thermal electrons that emit synchrotron radiation.

\begin{figure}
\centerline{ \epsfig{file=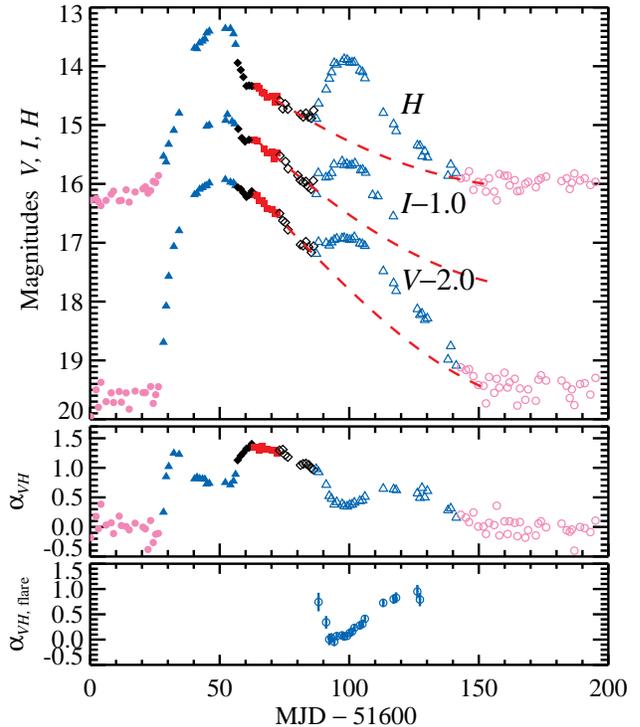, width=8cm}}
\caption{Light curve of \source\ in three filters $V$, $I$ and $H$ during the 2000 outburst. 
The $V$- and  $I$-magnitudes are shifted upwards by $2.0$ and $1.0$, respectively. 
Pink symbols correspond to the quiescent state, other symbols are the same as in Fig.~\ref{fig_xrays_lc}.
The red dashed lines show the evolution of the irradiated disc during the second flare (see Section~\ref{sec:cmd_nth}).
The middle panel show the power-law spectral index of the intrinsic spectrum $\alpha_{VH}$ corrected for extinction (with $A_V=5.0$)
determined from the $V$--$H$ colour using equation (\ref{eq:alpha_VH}).  
The lower panel shows the evolution of the spectral index of the flare component. 
}
\label{fig_oir_lc}
\end{figure}

\section{Light curves and extinction}
\label{sect:lc}

\subsection{Data}

The X-ray data on the 2000 outburst of \source\ 
covering 2.5--25 keV range from the Proportional Counter Array (PCA) spectrometer 
\citep{JMR06} on board the {\it Rossi X-ray Timing Explorer  (RXTE)}  
were analysed with the {\sc heasoft} package (version 6.15) and 
response matrices were generated using {\sc pcarsp} (11.7.1).  
The instrumental background of the PCA detectors was estimated with \verb|CM_bright_VLE| model. 
At the position of the source \source\ there is additional sky background from the so-called Galactic ridge \citep[see][ and references therein]{RSGCS06},  
which provides the flux at the level of $F_{\rm 3--20\,keV}\sim1.2\times 10^{-11}$~erg~s$^{-1}$~cm$^{-2}$
within field of view of PCA  ($\sim1$~deg$^2$). 
In order to account for this additional background, we have used  {\it RXTE}/PCA observations of \source\  in 2001 April when 
the source was already in quiescence.  
All the spectral data were fitted using {\sc xspec} 12.8.1g package \citep{Arn96}, assuming 1 per cent systematic uncertainty. 
To estimate fluxes from the source, we have fitted the spectral data with a standard for BHs  {\sc diskbb+powerlaw} model 
and corrected the model fluxes in the specific energy bands  by the ratio of the data to the model.
For a more detailed spectral modelling we also fitted the data with a hybrid Comptonization model  
{\sc compps} \citep{PS96} and a cutoff power-law model with Compton reflection {\sc pexrav} \citep{MZ95}. 
Interstellar absorption was taken into account using {\sc wabs} model with 
the  neutral hydrogen column density  of $N_{\rm H}=0.80 \times 10^{22}$~cm$^{-2}$ \citep{Miller03}.

The X-ray light curve is shown in Fig.~\ref{fig_xrays_lc} together with the evolution of the hardness ratio 
as well as with  the hardness-flux diagram. It is coloured according to the hardness ratio. 
The hard state is shown by blue triangles, while 
the  transitions from the hard to the soft state and back  are shown by black diamonds. 
The filled and open symbols correspond to the rising and decaying phases of the outburst, respectively. 
The soft state with the nearly constant hardness ratio is shown by red squares.

The OIR data from the 2000 outburst in  $V$, $I$ and $H$ filters have been presented by \citet{Jain01b}.  
To convert magnitudes to fluxes we use the  zero-points of 3636,  2416 and 1021~Jy 
and the effective wavelengths of 545, 798, and 1630~nm for  filters $V$, $I$ and $H$, respectively \citep{Bessell98}. 
We show the daily-averaged OIR light curves  in Fig.~\ref{fig_oir_lc}.
The colour code is the same as in Fig.~\ref{fig_xrays_lc},
with additional magenta points corresponding to the quiescent state of the source.
The fast rise--exponential decay morphology is accompanied here with flares, which are most prominent in the $H$ filter.

\subsection{Soft state and implications for the disc temperature}
\label{sect:ss_disc_temp}

\begin{table}
\caption{Parameters of the system adopted from \citet{Orosz11}.}
\begin{tabular}{lll}
\hline
Orbital period & $P_{\rm orb}$ & 37 \mbox{h} \\
Distance              & $D$        & 4.38 \mbox{kpc} \\
Black hole mass  & $M_{1}$ & 9.1$\msun$  \\
Companion mass & $M_{2}$ & 0.3$ \msun$  \\
Inclination & $i$ & 75\degr \\
Separation            & $a$        & 11.85$\rsun$ \\ 
Radius of the companion & $R_2$ & 1.75$\rsun$ \\ 
Roche lobe size & $R_{L,1}$ & 7.7$\rsun$ \\
Effective temperature of the companion & $T_{\rm eff}$ & 4475 K \\
\hline
\end{tabular}
\label{tab:parameters}
\end{table}

During the soft (and the following intermediate) state the OIR emission is likely originating in the irradiated accretion disc alone. 
This is supported by a simple exponential shape of the light curves in both the OIR and X-ray bands. 
This knowledge can be used to estimate the accretion disc temperature.
The first constraint relates the peak X-ray luminosity to the reprocessed optical flux.
The effective temperature at the outer radius of the irradiated disc dominating the OIR emission  is
$T_{\rm eff,irr}=[\eta (1-A) L_{\rm X}/(4\pi R_{\rm irr}^2 \sigma_{\rm SB}) ]^{1/4}$, 
where $A$ is the disc albedo and 
the factor $\eta\approx \frac{H}{R} \left(\frac{d\ln H}{d\ln R} - 1\right) $ gives the cosine between the normal to the 
outer disc and the direction to the central X-ray source \citep{FKR02}. 
The disc size $R_{\rm irr}$ is a fraction of the Roche lobe (\citealt{Eggl83}, see parameters in Table~\ref{tab:parameters}),
\begin{equation} 
R_{L,1}= a \frac{0.49 q^{2/3}}{0.6q^{2/3} + \ln (1+q^{1/3})}. 
\end{equation} 
For measured $a=11.85\rsun$ and $q=1/30$, we get $R_{L,1}=7.7\rsun$. 
The maximum disc size  limited by tidal forces is \citep{Warner95}  $R_{\rm irr}\lesssim 0.6a/(1+q) \approx 4.8 \times 10^{11}$~cm.

To estimate $\eta$, we take the disc half-opening angle of 12\degr\  \citep{deJong96}, i.e. $H/R\sim0.2$. 
The logarithmic derivative  $d\ln H/d\ln R - 1$ takes values  1/8  for the standard  and 2/7 for the irradiated discs  \citep{FKR02},
which we adopt in the following and get $\eta\approx 0.06$. 
A  typical albedo $A$ of mostly neutral material is below 0.5 even for a very hard power-law 
spectrum extending to 100 keV \citep{Basko74,MZ95}  and is expected to be below 0.1 
for the blackbody-like X-ray spectra with $kT_{\rm bb}\sim 1.5$~keV in the soft state.  
The bolometric luminosity at the peak of the 2000 outburst was $\approx 10^{38}$~\ergs\  (see Section~\ref{sec:oir_xray}). 
Because only photons with energy above 2~keV thermalize efficiently \citep{Sulei99},  
we use  $L_{>2{\rm keV}}\approx 5\times 10^{37}$~\ergs\ and finally get $T_{\rm eff,irr}\gtrsim 11\,000$~K. 
The temperature can be  lower if  the X-rays are so strongly anisotropic that the flux directed towards 
the outer disc  is significantly lower than that at the observed inclination of $i=75\degr$.
However, if the outer disc is inclined at a similar angle of $\sim80\degr$ \citep{deJong96}, practically no difference is expected. 
  
Further constraints are coming from the comparison of the decay rates in the X-ray and optical bands \citep{Endal76,vPMC95}.
The general idea is to compare the derivative of the observed flux over temperature to that of the known function -- 
a simple blackbody or the irradiated disc model spectrum.
Because the decay rate is an injective (i.e. one-to-one) function of temperature, we immediately obtain the absolute value of $T_{\rm eff}$.
 
If we ignore energy dissipation intrinsic to the disc (which is possible for high X-ray luminosities), 
the effective temperature of the irradiated disc varies with the X-ray luminosity as 
$T_{\rm eff}\propto L_{\rm X}^{1/4}$. 
Some deviations from this law are possible if the emission pattern is changing. 
Because the X-ray light curve shows some flares at the transition from the soft to the intermediate state, 
we have selected the soft-state segment of the data (red squares in Fig.~\ref{fig_xrays_lc}) 
and fitted the X-ray flux there with an exponential profile. 
We obtain the e-folding time of $\tau_{\rm X}=10.0\pm0.1$~d. 
This translates to the time of temperature decay $\tau_{\rm T}=4\tau_{\rm X}=40$~d, i.e.: 
\begin{equation} \label{eq:tefft_obs}
\partial \ln T_{\rm eff}/\partial t= -1 / 40 \ \mbox{d}^{-1}.  
\end{equation} 
We then fit the soft- and intermediate-state OIR light curves together with the values in the quiescence 
at MJD 51645--51650 with a constant  plus an exponentially decaying component. 
For $I$ filter, we adopt the constant $I=19$ in quiescence taken from the earlier observations \citep{Jain01a}.
The e-folding time-scale is then related to the derivative of the logarithm of the flux of the varying (disc) component: 
\begin{equation} \label{eq:bt_obs}
 \partial \ln F_{\lambda}/\partial t = -1 / \tau_{\lambda} \ \mbox{d}^{-1}.  
\end{equation}
We find $\tau_{\lambda}=22.6\pm0.8$, $26.0\pm1.0$ and $31.3\pm2.0$~d for  the $V, I$ and $H$ bands, respectively.
The decay of the blackbody flux due to the decreasing temperature can be computed as
\begin{equation} \label{eq:bt_theor}
\frac{ \partial \ln B_{\lambda}  } {\partial \ln T_{\rm eff} }
= \frac{ y } { 1-\exp(-y) },
\end{equation} 
where $y=hc/\lambda kT_{\rm eff}=1.44/(\lambda_{\mu} T_4)$, $\lambda_{\mu}$ is the wavelength in microns and $T_4=10^{-4}T_{\rm eff}$.
The value for the logarithmic derivative is then obtained by dividing equation~(\ref{eq:bt_obs}) by equation~(\ref{eq:tefft_obs}). 
Solving the resulting equation for $T_{\rm eff}$,  
we get the average effective soft- and intermediate-state temperatures of 20\,800, 19\,540 and 17\,500~K, 
with the errors on individual measurements much smaller than the spread between the values.
 
We checked the possibility of neglecting the constant flux in the OIR light-curves, which resulted in the decay times for the $V$, $I$ and $H$ bands
$\tau_\lambda=23.8\pm0.7$, $29.2\pm1.1$ and $39.2\pm2.0$~d, respectively.  
These values translate to the temperatures  23\,200 and 27\,400~K for $V$ and $I$ filters and for the $H$ filter 
temperature exceeds 200\,000K. 
If instead only the soft state light curve is fitted with the exponential, we get 
$\tau_\lambda=29.5\pm2.7$, $32.7\pm3.6$ and $41.9\pm5.4$~d, which 
correspond to $T_{\rm eff}$ in excess of 40\,000 K for $V$ and $I$ filters and again  no solution is possible for the $H$ filter. 

We also checked, whether the irradiated disc model gives any improvement compared to the simple blackbody. 
In this case equation~(\ref{eq:bt_theor}) can be rewritten as
\begin{equation} \label{eq:bt_theor_irr}
\frac{ \partial \ln {F_{\rm irr, \lambda}} }{ \partial \ln T_{\rm out} }=
 y_{\rm o}  \displaystyle 
\frac{\displaystyle \int_{r_{\rm io}}^1 \frac{\exp(y_{\rm o}x^\beta)}{[\exp(y_{\rm o}x^\beta)-1]^2} x \rmd x }{\displaystyle \int_{r_{\rm io}}^1 \frac{x\ \rmd x}{\exp(y_{\rm o}x^\beta)-1}},  
\end{equation} 
where $\beta$ is the power-law index of the radial temperature dependence 
$T(R)\propto R^{-\beta}$, $r_{\rm io}=R_{\rm in}/R_{\rm out}$ is the ratio of the inner to outer disc radius and $y_{\rm o}=hc/\lambda kT_{\rm out}$. 
In Rayleigh--Jeans regime $y_{\rm o}\ll 1$, the logarithmic derivative takes the minimum value of $(1-\beta/2)/(1-\beta)$.  
We see that for realistic $\beta=3/7$--$1/2$ \citep{Cun76,FKR02}, 
the model minimum derivatives of 1.4--1.5 are larger than 
the observed for the $H$ filter values $40/\tau_\lambda=$0.95--1.3, thus no solution is possible. 
Hence, the simple irradiated disc does not give a good description to the shape of the light curve.  

We conclude here that from the simple estimate of the maximum disc size and peak X-ray luminosity, 
the outer disc temperature has to be above 11\,000 K in the  soft and intermediate states, while from the 
light curve behaviour we get  $T_{\rm eff}\gtrsim18\,000$~K. 
Thus we conservatively assume that  $T_{\rm eff}\gtrsim15\,000$~K during the peak of the 2000 outburst.

\subsection{Extinction towards \source\ and spectral index -- colour relation}
\label{sec:av_colour}

Optical extinction towards \source\ can be estimated from the hydrogen column density.  
The {\sc ftools} routine NH \citep{Dickey90} gives  $N_{\rm H}=0.9\times 10^{22}$~cm$^{-2}$. 
\citet{TCK01} has obtained $N_{\rm H}=(0.85^{+0.22}_{-0.24})\times 10^{22}$~cm$^{-2}$   
using  {\it Chandra} observations in the end of the 2000 outburst. 
Similar result, $N_{\rm H}=(0.88^{+0.12}_{-0.09})\times 10^{22}$~cm$^{-2}$, 
was obtained by \citet{Corbel06} again from the {\it Chandra} observations  in a quiescent state. 
The most accurate measurement of $N_{\rm H}=(0.80\pm0.04)\times 10^{22}$~cm$^{-2}$, 
which we use for X-ray data analysis, is from the {\it Chandra} observations during the peak of the 2000 outburst  \citep{Miller03}. 
Thus, all measurements are consistent with $N_{\rm H}$ lying in the interval $(0.75-1)\times 10^{22}$~cm$^{-2}$. 
However, these results are inconsistent with the earlier estimate of the extinction 
$A_V=2.2\pm0.3$ based on interstellar optical absorption lines \citep{SFCT99}. 
The later measurements and modelling of \citet{Orosz11} imply $A_V\approx 4.75$. 
The \citet{PS95} relation, $A_V=5.59\ N_{\rm H}/10^{22}$, gives $A_V$=4.2--5.6.  

\begin{figure}
\centerline{ \epsfig{file=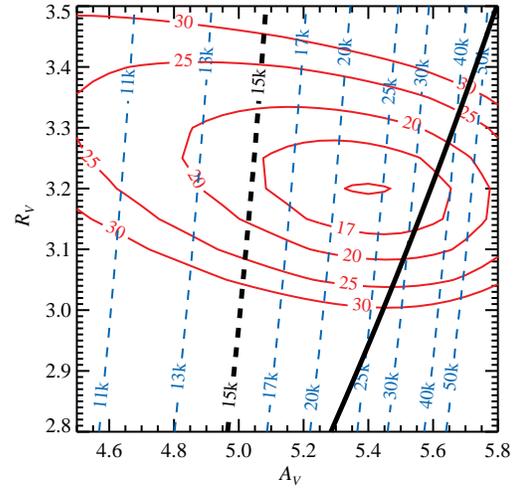, width=6.5cm}}
\caption{Contours of $\chi^2$ (red solid curves) for the blackbody fit to the OIR data  
in the soft and intermediate state (MJD  51663--51676) at the plane $A_V$--$R_V$. 
The dashed contour lines show the average  temperature  
in these states varying from about 10\ 000~K for $A_V=4.5$ to 30\ 000~K for $A_V=5.5$. 
The thick dashed black line marks the position of the average $T_{\rm eff}=15\,000$~K, 
thus the region to the left of this line is forbidden. 
The region to the right of the thick  solid black line corresponds to $\alpha_{VI}>2$, i.e. 
harder than the Rayleigh--Jean and therefore is forbidden.  
}
\label{fig_chi2_av}
\end{figure}

Another way of estimating $A_V$ is to  use constraints on the disc temperature from Section~\ref{sect:ss_disc_temp}. 
We have converted the OIR magnitudes to fluxes  and 
fitted the three-point spectra obtained  in the soft and intermediate states 
with the blackbody disc of the constant radius $R_{\rm irr}$ and varying temperature: 
\begin{equation} \label{eq:fluxbb}
 F_{\nu}(T_{\rm eff}) =  \cos{i} \  B_{\nu}(T_{\rm eff})\ \pi R_{\rm irr}^2/ D^2.
\end{equation}
The extinction $A_V$ was allowed to vary. 
We also checked how the shape of the extinction curve affects the results by varying $R_V$. 
We used the extinction law of  \citet{Cardelli89}  corrected by \citet{odonnell94}. 
All together we fitted 13 one-day-averaged spectra (39 points) using 14 fitting parameters ($R_{\rm irr}$ and 
13 temperatures) with 25 d.o.f.  
The contours of $\chi^2$ on the plane $R_V$--$A_V$ are shown in Fig.~\ref{fig_chi2_av}. 
We see that the best fit with $\chi^2=14.9$ is achieved for $R_V=3.2$ and $A_V=5.4$. 
The effective disc radius is $R_{\rm irr}=2.85\times 10^{11}$~cm, which  
is 40 per cent smaller than  the maximal possible disc size limited by tidal forces. 
We should remember, however, that the actual  emission area is likely a ring, not a circular disc, therefore
the actual disc size is larger. 
We also plot at the same plane the average temperature from the best-fitting models. 
The lower limit on the typical temperature  $T_{\rm eff}>15\,000$~K obtained in Section~\ref{sect:ss_disc_temp} 
can be now transformed to a lower limit of $A_V>5.0$. 
These results are almost independent of $R_V$.

Two more constraints can be obtained from the extreme colours shown by \source. 
For that it is useful to get the relations between colours and the corresponding 
intrinsic (without absorption) spectral indices ${\alpha}$ of the  power-law spectrum $F_{\nu}\propto \nu^{\alpha}$. 
The apparent magnitude in any filter is defined as 
\begin{equation}
m_{\nu} = -2.5\log F_{\nu}/F_{\nu,0} + A_{\nu} ,  
\end{equation}
where $A_{\nu}$ is the extinction, $F_{\nu}$ is the intrinsic flux without absorption and $F_{\nu,0} $ is the zero-point.
The slope between bands $i$ and $j$ can be computed as 
\begin{equation}
\alpha_{ij} \equiv  \frac{\displaystyle \log \left(\frac{F_i}{F_j}\right)}{ \displaystyle \log \left(\frac{\nu_{{\rm eff}, i}}{\nu_{{\rm eff}, j}}\right)} = 
 \frac{m_i - m_j - A_i +A_j - 2.5 \displaystyle \log\left(\frac{F_{i,0}}{F_{j,0}}\right) }
 {2.5\displaystyle  \log \left(\frac{\lambda_{{\rm eff}, i}}{\lambda_{{\rm eff}, j}}\right)} ,
\end{equation}
where $\lambda_{\rm eff}$ are the effective wavelengths of the corresponding filters.  
The slopes between the considered bands $V$, $I$ and $H$ are then:\footnote{These transformation 
laws correct the erroneous expressions presented in \citet{RMDF11},
which have wrong scalings and give index $\alpha$ of about 1.0 too small.
They have used the wavelength and the zero-point of the $J$ filter 
instead of those for the $H$ filter in the equation similar to our equation (\ref{eq:alpha_VH}), 
while in the equation corresponding to our equation (\ref{eq:alpha_IH}), 
they have taken the values for the Johnson $I$ filter instead of the Kron--Cousins $I$ filter 
 (Dipankar Maitra and David Russell, private communication).}  
\begin{eqnarray} \label{eq:alpha_VH}
\alpha_{VH} & =& 1.16 + 0.69\,A_V - 0.84\ (V-H) , \\ 
 \label{eq:alpha_IH}
 \alpha_{IH} & =& 1.21 + 0.55\,A_V - 1.29\ (I-H), \\ 
  \label{eq:alpha_VI}
 \alpha_{VI} & =& 1.07  + 0.94\, A_V - 2.42\ (V-I) , 
\end{eqnarray}
where we used the ratios $A_I/A_V=0.61$ and $A_H/A_V=0.185$ obtained for $R_V=3.2$.

The  bluest spectra observed from 
\source\  during a brighter outburst in 1998 \citep{Jain99}, which had $V-I=1.8$ (for $V=16.8$ and $I=15.0$), 
have to be softer than the Rayleigh--Jean tail of the blackbody, i.e. $\alpha_{VI}<2$.
Using  equation (\ref{eq:alpha_VI}), we now immediately get a firm upper limit on $A_V\lesssim 5.6+0.75\,(R_V-3.2)$ 
shown by a thick black solid line in Fig.~\ref{fig_chi2_av}. 

On the other hand, the reddest spectra in the beginning and the end of the 2000 outburst still consistent with the blackbody have $V-H\approx 5.3$. 
According to the disc instability models \citep{DHL01,Lasota01,FKR02}, 
for the outburst to start, the outer disc temperature has to be above the hydrogen ionization temperature of about 6000 K. 
The blackbody of this temperature has $\alpha_{VH}\approx0.1$. 
Thus, we get the lower limit from equation (\ref{eq:alpha_VH}): $A_V\gtrsim4.9$, which 
is nearly identical to the constraint we get from the soft-state spectra in Section~\ref{sect:ss_disc_temp}.

We conclude that a realistic range of the extinction is $A_V=$5.0--5.6.
In the further discussion, we adopt $A_V=5.0$ (and $R_V=3.2$), with 
extinction in other filters is then $A_I=3.05$ and $A_H=0.92$. 
Note that if we assume larger $A_V$, the typical spectral indices would then be larger 
than given by equations (\ref{eq:alpha_VH})--(\ref{eq:alpha_VI}).
This does not change qualitatively the behaviour of the source and does not 
affect any of the conclusions.

\section{Spectral properties} 
\label{sect:spectra}

\subsection{$V$ versus $V-H$ diagram}
\label{sect:vvh}

\begin{figure}
\centerline{ \epsfig{file=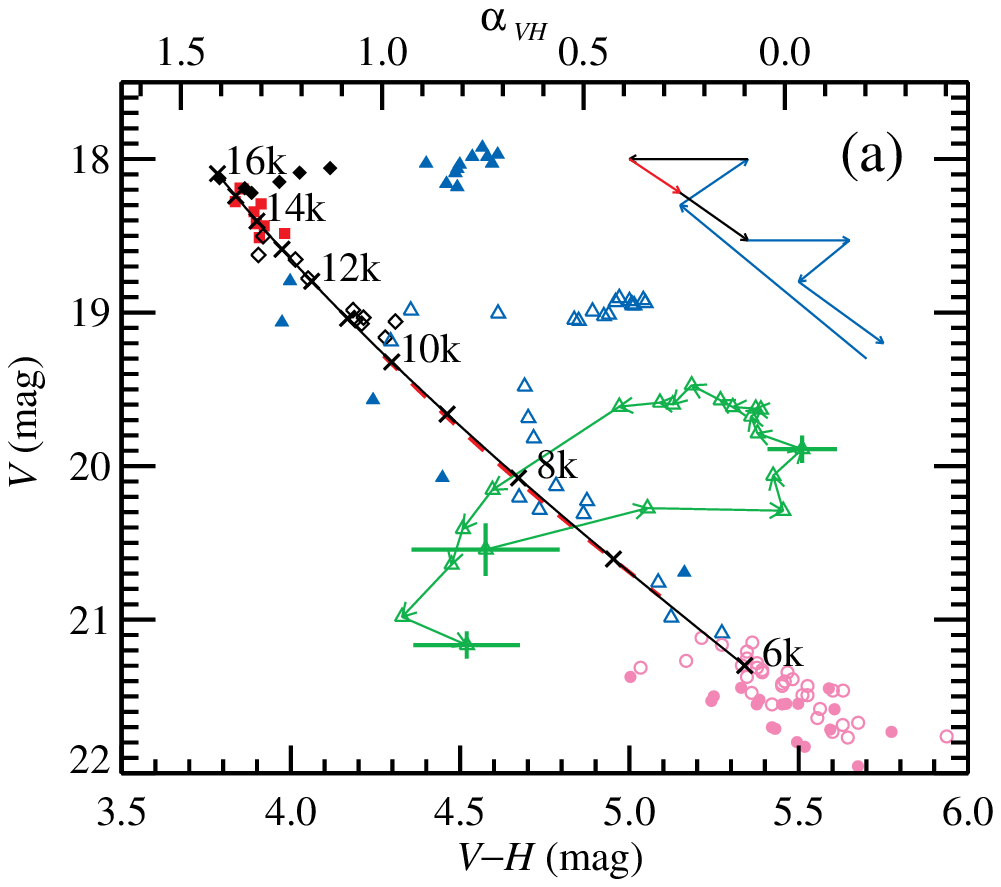, width=6.5cm} }
\centerline{  \epsfig{file=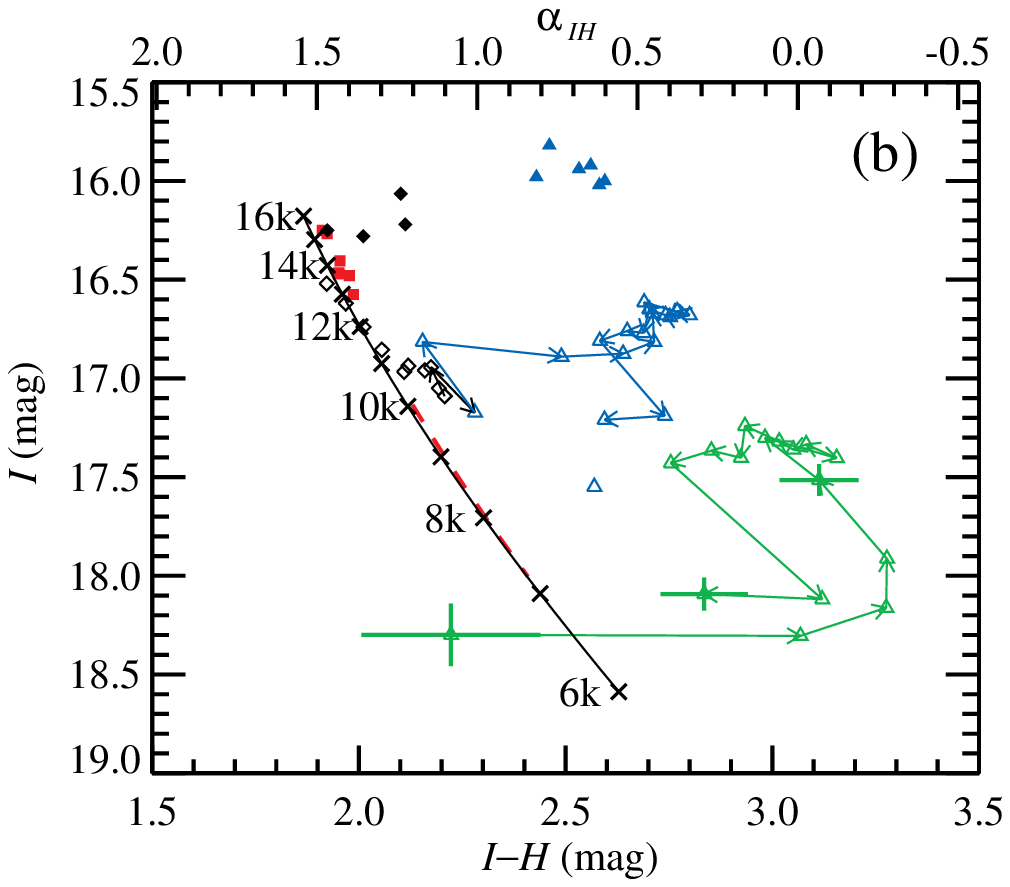, width=6.5cm} }
\centerline{  \epsfig{file=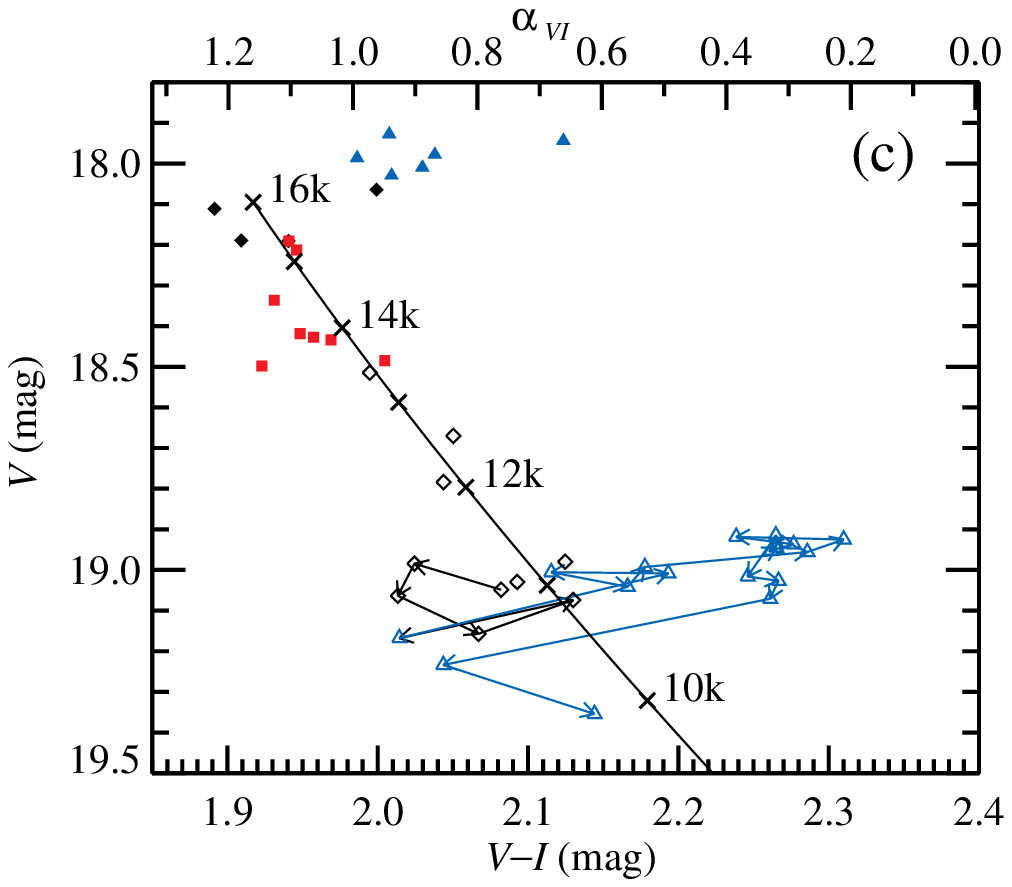, width=6.5cm} }
\caption{The observed  (a)  $V$ versus $V-H$,  (b) $I$ versus $I-H$  and (c) $V$ versus $V-I$   CMDs.  
The coloured track in the upper-right corner of panel (a)  illustrates the path \source\ follows at the diagram. 
The black solid lines represent the theoretical curves for the blackbody disc of radius $R_{\rm irr}=2.85\times10^{11}$ cm 
inclined at $i=75\degr$ at distance of 4.38~kpc \citep{Orosz11} of different temperatures (in units of thousands of Kelvin, marked next to the line). 
The model magnitudes were reddened following the extinction law of \citet{Cardelli89} and \citet{odonnell94} with  $A_V=5$ and $R_V=3.2$. 
The upper $x$-axes show the intrinsic spectral indices in the corresponding wavelength bands given by equations (\ref{eq:alpha_VH})--(\ref{eq:alpha_VI}).   
The red dashed line represents the disc component extrapolated to the time of the flare in the hard state (see Fig.~\ref{fig_oir_lc}). 
The blue and black arrows connect points in the end of intermediate and beginning of the hard states.
The green track shows the path made by the flare component during the hard state starting from MJD~51688 
(see Section~\ref{sec:cmd_nth}).   
}
\label{fig_oir_cmd}
\end{figure}

Fig.~\ref{fig_oir_cmd}(a) represents the $V$ versus $V-H$ colour--magnitude diagram (CMD).
The symbols correspond to the outburst stages identified from the X-ray hardness ratio (Fig.~\ref{fig_xrays_lc}). 
The path the source makes on the diagram is illustrated in the upper-right corner  with arrows. 
The black line corresponds to theoretical colour--magnitude relation for a blackbody of different temperatures with the disc radius 
determined from the best fit to the soft- and intermediate-state spectra (assuming $A_V=5$). 
The upper $x$-axis at the CMD shows the slope $\alpha$ of the intrinsic spectrum between the corresponding bands 
converted from the colour using equation  (\ref{eq:alpha_VH}). 

At the beginning of the outburst their evolution can be well described by a blackbody of increasing temperature, until the colour $V-H\sim4.0$ is achieved.
After that, the source becomes significantly redder than a blackbody of the corresponding $V$ magnitude.
This behaviour can be interpreted as an appearance of an additional, non-thermal component.
At the X-ray hard-to-soft transition, the source returns to the blackbody track at almost constant $V$ magnitude 
(horizontal track in the CMD), indicating the quenching of the  additional component around MJD~51659. 
In the soft and in the beginning of the following intermediate state, 
the OIR colours are well described by a blackbody of decreasing temperature. 
During the reverse transition,  
we observe re-appearance of the red component around MJD~51683, again as a horizontal track in the CMD. 
Then the source slowly decays towards the quiescence along the blackbody track of 
the same normalization as during the soft state, which is, however, slightly larger  than that 
at the rising phase of the outburst.
The peculiar fast changes in the $H$ band at almost constant $V$ band put constraints on the possible origin of non-thermal emission.

We also have checked, whether additional contribution from the secondary star affects the overall shape of the magnitude--colour relation.
We used the atmosphere templates of \citet{Castelli04} for a K3 star with $\log g=3.5$, $T_{\rm eff}=4500$~K 
and radius of $1.75\rsun$ \citep{Orosz11}. 
The secondary has  a minor effect on the CMD, except for quiescence, where the model becomes slightly redder than the data.  
If, however, we assume a larger extinction, e.g. $A_V=5.2$, the blackbody temperature 
corresponding to the soft state increases to 20\,000--25\,000 K and the total theoretical CMD, then well 
describes the data even in the quiescent state. 
Because we are not interested in this state (and for the sake of simplicity of the model), we further do not account for the secondary 
contribution.

\begin{figure}
\vspace{0.5cm}
\centerline{ \epsfig{file=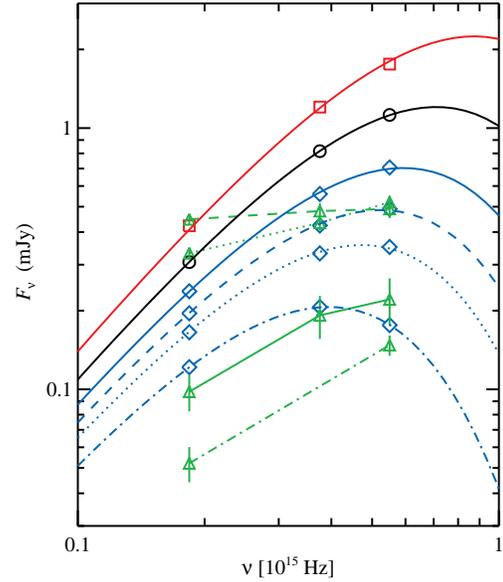, width=6.5cm}}
\caption{Examples of spectra (corrected  for extinction) and the best-fitting blackbody model of  
constant emission area. 
Data taken in the soft state on MJD 51665  and in the intermediate state on MJD 51676 are 
shown by red squares and black circles,  respectively. 
In the hard state, the disc fluxes on MJD 51688, 51697, 51706 and 51726 
obtained by interpolation are shown by blue diamonds and the flare component by green triangles.   
The best-fitting blackbody models are shown by solid, dashed, dotted, dot--dashed blue lines, respectively. 
The errors are not shown for most data points, because they are smaller than the size of the symbols. 
}
\label{fig_spe_all}
\end{figure}

\subsection{$I$ versus $I-H$ diagram}
\label{sect:iih}

We also plot the $I$ versus $I-H$ CMD in Fig.~\ref{fig_oir_cmd}(b).
The behaviour of the source in this diagram is very similar to that in the $V$ versus $V-H$ CMD. 
The blackbody track (black solid line) for the same parameters as in Fig.~\ref{fig_oir_cmd}(a) provides a good fit to the data
from MJD~51660 at the end of the hard-to-soft (HS) transition, 
in the soft state and in the beginning of the soft-to-hard (SH) transition until about MJD~51683.   
In the middle of the SH transition, 
 the source becomes redder indicating the presence of the additional component.   
The X-ray hardness during the moments of appearance/disappearance of this component 
was nearly the same (see Fig.~\ref{fig_xrays_lc}c), 
corresponding to the X-ray spectrum with photon index $\Gamma\approx 1.9$ 
and reflection amplitude $\Omega/(2\pi)\approx 1$.
 
An interesting additional detail is a hook-like evolution (marked by blue arrows) 
at the end of the hard state, before the flare in the $H$ filter.  
This feature arises because several data points at the beginning of the hard state are  redder at lower $I$ flux compared to the 
data points a few days later, 
which makes  the spectrum bluer (larger $\alpha_{IH}$) at a higher $I $flux. 
This behaviour cannot be   
an artefact of interpolation between non-simultaneous points in the $H$ filter 
and can be interpreted as the flare starting first in the $I$-band, then proceeding to the $H$ band.
This is consistent with the shape of the flare spectra (see Section~\ref{sec:cmd_nth} and Fig.~\ref{fig_spe_all}).

The spectral index  $\alpha_{IH}$  (upper $x$-axis)  
at the peak of the flare (when the spectrum is softest in the hard state) 
is identical to $\alpha_{VH}$,  implying a single power-law component going through all three filters.
We note that taking larger values of extinction results in a larger $\alpha$ (harder spectra), 
but the effect is larger for $\alpha_{VH}$ than for $\alpha_{IH}$, 
thus the spectrum would no longer be described by a single power law.

\subsection{$V$ versus $V-I$ diagram}
\label{sect:vvi}

The general evolution of the source at the $V$ versus $V-I$  CMD shown in  Fig.~\ref{fig_oir_cmd}(c)  is very similar to those 
for other filters as described above. However, at this diagram the appearance of the additional non-thermal component 
is easier to see. Around MJD~51683, in the middle of the intermediate state, 
the spectrum becomes bluer, which is reflected in motion of the source at $V=19$ to the left 
from the blackbody model line by $\Delta(V-I)\approx 0.1$ (shown by black arrows). 
This implies that   the additional component has a spectrum harder than the blackbody spectrum at this moment of $\alpha_{VI}\approx 0.7$. 
The spectrum becomes redder that the corresponding blackbody at MJD~51688. 
Only at this moment, the flare starts to become easily visible in the light curves.

\subsection{Flares and the non-thermal component}
\label{sec:cmd_nth}

The underlying behaviour of light curves in $V$, $I$ and $H$ filters is naturally explained with the evolving temperature of the accretion disc, 
which responds to changes in the X-ray luminosity (i.e., due to changes of the reprocessing flux). 
The CMDs as well as the spectra at the soft and the following intermediate states 
suggest that the OIR emission comes from the irradiated accretion disc that  is modelled here as a blackbody 
(see Fig.~\ref{fig_spe_all} and Sections~\ref{sect:vvh} and \ref{sect:iih}).
The flares that occur in the intermediate/hard states can be  interpreted as the appearance of an additional, non-thermal component. 
The first flare starts at the rising phase of the outburst and 
it is therefore impossible to separate it from the underlying disc emission. 
On the other hand, the second flare occurs at the exponentially decaying stage 
and it is possible to subtract the disc emission in order to isolate the flaring component.

To obtain the disc emission at the time of the second flare, 
we use the same model as in Section~\ref{sect:ss_disc_temp}, i.e. 
we fit the light curves in every filter with the constant plus exponential that 
represents  the irradiated disc. The fit is applied to the times MJD 51663--51683 
for all filters and MJD 51750--51755 for $V$ and $H$ filters.
The first interval correspond to the soft and intermediate states where there no signatures of the 
additional component, while the second interval correspond to the quiescent state. 
Unfortunately, no $I$ filter data exist after  MJD 51717 and 
similarly to Section~\ref{sect:ss_disc_temp}, we fix the constant corresponding to $I=19$. 
The best-fitting models are shown in Fig.~\ref{fig_oir_lc} by dashed red curves. 
After the fitting, we check that the magnitudes and colours of the extrapolated disc component are consistent with the 
blackbody of the same size as before the flare (compare red dashed lines in Fig.~\ref{fig_oir_cmd} to the black solid lines representing the
blackbody model). 
We see that the typical accuracy of the magnitudes of the interpolated blackbody component is better than 0.03. 

By subtracting the blackbody flux at every moment, we extract the flaring non-thermal component.
The evolution of colours and spectral indices of the flare is shown in Fig.~\ref{fig_oir_cmd} by green triangles connected by 
arrows. The typical errors are shown by crosses for a few points. 
From Fig.~\ref{fig_oir_cmd}(a) we see that the flare starts with the hard index $\alpha_{VH}=0.75\pm0.20$, 
then softens down to $\alpha_{VH}\approx0$ and then hardens again to $\alpha_{VH}=0.7-0.8$. 
The time evolution of $\alpha_{VH}$ of the total spectrum and of the flaring component 
is shown in the middle and lower panels of Fig.~\ref{fig_oir_lc}. 

The flare path in $I$ and $H$ filters is seen in  Fig.~\ref{fig_oir_cmd}(b). 
The evolution is similar to the $V-H$ colours, but  the spectrum is even harder at the start of the flare 
with $\alpha_{IH}=1.1\pm0.3$, implying that the spectrum hardens towards longer wavelengths 
(see green solid line in Fig.~\ref{fig_spe_all}).  
At the peak of the flare  $\alpha_{IH}\approx +0.2$ nearly identical to $\alpha_{VH}$ implying that 
the spectrum in three filters is close to a power law of the same index. 
At the end of the flare, the spectrum in $I$ and $H$ band is less  reliable, because of the 
growing error in the disc flux in $I$ filter. 

The flare becomes visible in the light curves only in the beginning of the 
hard state at MJD~51688, but substantial deviations from the blackbody spectrum occur already at MJD~51683 
as can be seen from the $V-I$ colour evolution shown in Fig.~\ref{fig_oir_cmd}(c).  
In the end of intermediate state, during MJD~51683--51687 
the total spectrum is bluer than the blackbody with $\alpha_{VI}\approx 0.7$ and reaches $\alpha_{VI}\approx 1$. 
This  implies that the slope of the flare spectrum between $I$ and $V$ filters is substantially 
larger than unity, because the flare contribution to the total flux  at this moment  is  still rather low. 

Fig.~\ref{fig_spe_all} demonstrates the evolution of the disc  and the flare spectra during the outburst. 
We see that the  flare spectrum is a broken power law just at the start of the flare, but 
is consistent later with a simple power law. 
Assuming a larger extinction, e.g. $A_V=5.2$, results in $\alpha$ larger by about 0.2, 
but the qualitative behaviour remains the same. 
We note here  that our conclusions on the evolution of the flare spectrum are based on the 
assumption that the blackbody normalization did not change during the decaying phase of the outburst. 
This is supported by the absence of significant variations in the normalization from the soft state to the 
end of the hard state as well as the quiescence.  
 
\begin{figure*}
\centerline{\epsfig{file=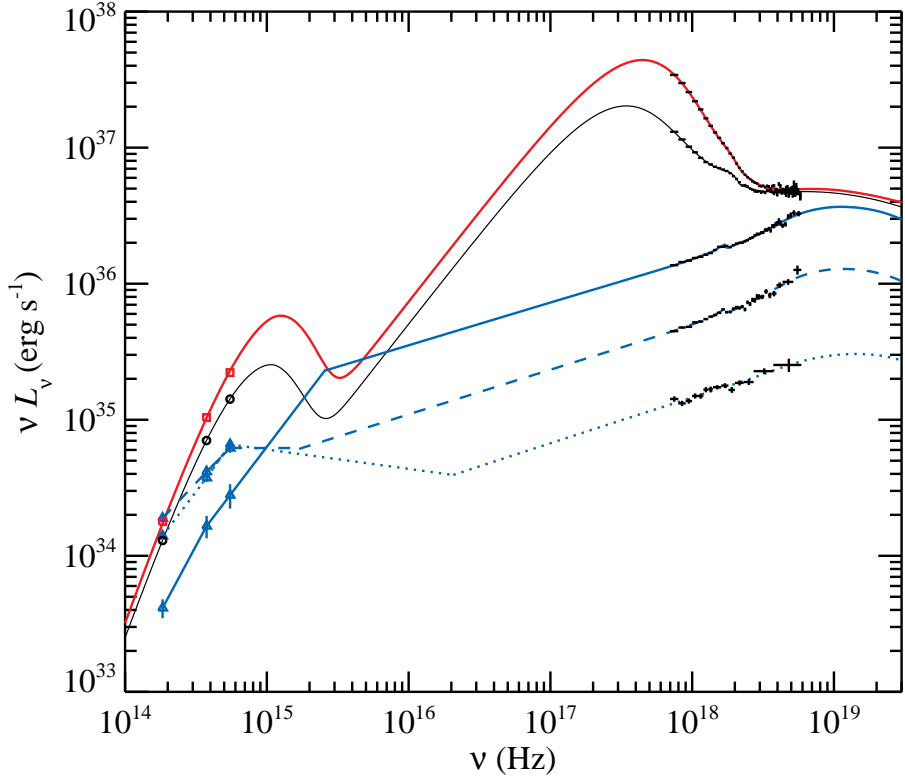, width=12cm}}
\caption{
Broad-band OIR to the X-ray spectral energy distribution of  \source. 
The black crosses are the {\it RXTE}/PCA data. 
Spectra are plotted for the same dates as in Fig.~\ref{fig_spe_all}, 
excluding the last one, which has very unreliable X-ray spectrum because of the 
strong background due to the Galactic ridge emission.  
The soft state data at MJD~51665  and the model are shown by red squares and the curve,  
the intermediate state at  MJD~51676 (black symbols), and the 
hard-state data  at MJD~51688, 51697, and 51706  by  blue triangles and by 
solid, dashed and dotted  curves, respectively. 
The blackbody component from the irradiated disc is shown in the OIR for the  soft and intermediate  states, 
while for the hard state only the flare component (same as in Fig.~\ref{fig_spe_all}) is shown. 
The hard-state OIR model spectra are somewhat arbitrarily connected to the power law extrapolated from the X-ray data. 
}
\label{fig_sed_broad_nuFnu}
\end{figure*}

\subsection{OIR -- X-ray relation}
\label{sec:oir_xray}

Once we have separated the non-thermal component in the OIR band, it is 
worth looking at a larger picture by understanding the relation between the OIR 
flare spectrum and the X-rays. Fig.~\ref{fig_sed_broad_nuFnu} shows 
the broad-band OIR/X-ray spectra of \source\ taken at different states. 
Similar X-ray data from the 2000 outburst of \source\ 
were presented before by \citet{Yuan07} and \citet{xue08}. 

The soft state X-ray spectrum (taken on MJD~51665) is dominated by a strong thermal component 
from the optically thick accretion disc. There is a high-energy tail above 20 keV. 
These non-thermal tails are well explained by non-thermal Comptonization in the hot corona 
above the disc \citep{PC98,Gier99,ZG04}. 
The thermally-looking component itself cannot be explained by a simple standard disc model, but 
requires a contribution from  thermal Comptonization. 
The whole X-ray spectrum thus can be fitted with a hybrid Comptonization model  such 
as {\sc eqpair} \citep{Coppi99} or {\sc compps} \citep{PS96} models in {\sc xspec}. Here we used the latter. 
We do not discuss here the best-fitting  parameters of the model, just because the aim of the fitting is 
only to show the shape of the spectrum, which is rather independent of the  model choice.  
The extrapolation of the standard cold accretion disc  {\sc diskbb} spectrum to lower energies 
reveals that it cannot contribute more than a few percent to the observed OIR flux, which is dominated by the reprocessed 
radiation from the outer disc with the total (unabsorbed) luminosity 
of $L_{\rm repr}\approx 7.3\times 10^{35}$~\ergs.  
This has to be compared to the total (unabsorbed) X-ray luminosity in the 0.01--1000 keV band 
of $L_{\rm X}\approx 1.2\times 10^{38}$~\ergs. 
Assuming the same angular dependence of radiation  for both components, 
the  ratio of luminosities gives the total reprocessing efficiency of $\epsilon_{\rm repr}\approx 6\times 10^{-3}$. 
The intermediate-state spectrum (on MJD~51676) is nearly identical to the soft-state spectrum, with a slightly lower temperatures 
and the X-ray and outer disc luminosities of $L_{\rm X}\approx 7.1\times 10^{37}$~\ergs\ and $L_{\rm repr}\approx 3.2\times 10^{35}$~\ergs, respectively, 
giving nearly the same reprocessing efficiency of $\epsilon_{\rm repr}\approx 4.5\times 10^{-3}$.

\begin{figure*}
\epsfig{file=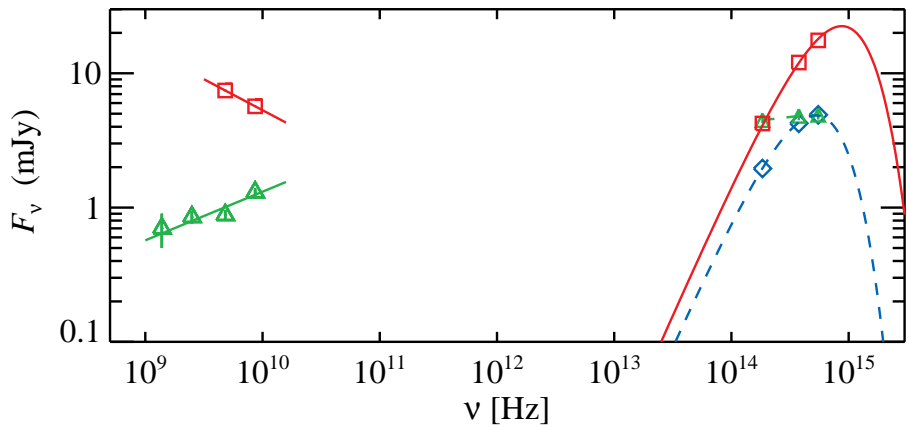, width=12cm}
\caption{
Broad-band radio to optical spectral energy distribution of  \source\ at MJD~51665 (red squares) and 
51697 (green triangles  and  blue diamonds).
The radio data are taken from \citet{Corbel01} and the OIR data from \citet{Jain01b}. 
We plot the blackbody with temperature and normalization as inferred from the light curves and the CMD.
The additional emission is due to the non-thermal component (shown with green symbols). 
In the soft state, the radio jet does not contribute at all to the OIR band, and in the hard state 
the flare OIR spectrum has a different slope $\alpha\approx0$ that does not lie on the continuation 
of the radio power law with $\alpha_{\rm radio}\approx0.36$. 
}
\label{fig_sed_radio_oir}
\end{figure*}

In the hard state, we  used the cutoff power-law model with Compton reflection ({\sc pexrav} model 
in {\sc xspec}; \citealt{MZ95}) for modelling and estimating the luminosities, because 
there are no any signatures of the thermal emission from the standard accretion disc in the X-ray band. 
We assumed a cutoff at 200 keV which is typical for a BH X-ray binary. 
During the three considered days (MJD 51688, 51697 and 51706), 
the spectral shape varied very little, with the power-law photon index  being nearly identical  at $\Gamma \approx1.65$ 
(i.e. $\alpha\approx-0.65$) and the reflection amplitude $\Omega/(2\pi)$ decreasing from about unity to 0.3. 
The X-ray luminosity took the values of $L_{\rm X} \approx (16, 5.6, 1.5)\times 10^{36}$~\ergs. 
The X-ray spectrum taken on MJD~51697 extrapolated to the OIR band (see blue lines in Fig.~\ref{fig_sed_broad_nuFnu}, 
where we plotted only the flare spectrum in the OIR) matches rather well with the OIR power-law. 
At the beginning of the hard state, at MJD~51688 
(solid blue line), the extrapolation of the X-ray power law is an order-of-magnitude above the OIR flux. 
On the other hand, later in the hard state, on MJD~51706, the X-rays underpredict the OIR emission.

In addition to the flare component, the emission from the irradiated disc (not shown in Fig.~\ref{fig_sed_broad_nuFnu}) 
was also steadily declining with $L_{\rm repr}\approx  (16, 9.5, 6.3)\times 10^{34}$~\ergs\ at the same three dates. 
This gives the reprocessing efficiency of $\epsilon_{\rm repr}\approx (1, 1.7, 4.2)\times 10^{-2}$. 
Thus we see  a clear trend of increasing $\epsilon_{\rm repr}$ with decreasing X-ray luminosity. 
This trend has also been reported for other BH transients \citep[e.g.][]{GDP09}. 
There could be at least two reason for such a change. First, the geometry of the accretion disc 
and therefore the emission pattern changes during the SH  transition. 
In the soft state, the emission from optically thick standard disc is beamed perpendicular to its 
plane, hence the outer disc sees less radiation than the observer at 75\degr\ inclination. 
In the hard state, however, the emission from optically thin hot disc  is much more isotropic \citep[see e.g.][]{VPI13}. 
Secondly, the reprocessing efficiency is higher for harder spectra, because 
the soft X-rays are absorbed in the very surface layers producing mostly UV lines and recombination continua 
(\citealt{Sulei99}; see also discussion in \citealt{GDP09}). 
We can account for the second effect by dividing $L_{\rm repr}$ by the luminosity above 2 keV, 
getting the  reprocessing efficiency of  $\epsilon_{\rm repr}\approx 1.2 \times 10^{-2}$  and $0.8 \times 10^{-2}$ 
in the soft and intermediate states, respectively, and  $\epsilon_{\rm repr}\approx (1.2, 2.1, 5.2)\times 10^{-2}$ in the hard state. 
This reprocessing efficiency is still rather modest because for the disc half-opening angle of 12\degr\ \citep{deJong96} 
we expect $\epsilon_{\rm repr}\lesssim 0.15$ assuming albedo of 0.3. 
The observed reprocessing efficiency can actually be even lower, because at low fluxes a significant fraction 
of the OIR radiation can be produced by the (ignored here) secondary star  
and/or the bulge located at the impact point of the accretion stream.

\section{Origin of non-thermal component}
\label{sect:origin}

\subsection{Jet?}

The non-thermal OIR component seen in \source\ was attributed to the jet synchrotron emission in a number of works 
\citep{Corbel01,RMDM10,RMDF11}.
The aforementioned works used the optical data, originally published in \citet{Jain01b}, the same as used in this work.
\citet{Corbel01} obtained optical $V-I$ spectral index $\alpha=-2.6$ (assuming $A_V=2.2$) on 2000 June 1 (MJD 51697), 
in contrast to the obtained by us the flare spectral index $\alpha\approx 0$ (for  $A_V=5.0$).
The source of discrepancy is the extinction value, poorly known at that time.
 
The studies by \citet{RMDM10,RMDF11} present the separation of the non-thermal emission using the light-curve fitting method.
They claimed that the spectral indices of the non-thermal component are in the range $-1.5\lesssim\alpha\lesssim-0.5$, 
again suggesting the non-thermal optically thin jet emission.
However, these works suffer from the major errors in formulae for 
transformation of the colours to indices (see Section~\ref{sec:av_colour}). 
Furthermore, their fits to the disc light curve in $V$ and $I$ filters are significantly above the data points 
just before the flare (see fig. 2 in \citealt{RMDM10})  leading to overestimation of the disc contribution 
and to over-subtraction of the flux in those filters resulting in a much too soft flare spectrum.
These flawed fits have greatly affected the first few points of the flare, where the non-thermal $V$ flux was small, 
hence only spectral hardening during the flare was detected, 
while we obtain that the flare starts with hard spectrum with $\alpha\approx +0.75$, 
which softens to $\alpha\approx0$ and then hardens again (see Fig.~\ref{fig_oir_cmd}). 
Such behaviour is clearly inconsistent with the optically thin jet spectrum.   
The jet radio emission was indeed optically thin in the soft state,  with $\alpha_{\rm radio}= -0.46\pm0.03$ (Fig.~\ref{fig_sed_radio_oir}), 
but it does not contribute at all to the OIR spectrum, which was consistent with the blackbody.

On the other hand, the radio emission in the  hard state close to the peak of  the second flare, on MJD~51697.14 
was optically thick  with  $\alpha_{\rm radio}= 0.36\pm0.09$ \citep{Corbel01}.  
However, the OIR flare spectrum has a different slope and does not lie on the continuation of the radio spectrum 
(see green points in Fig.~\ref{fig_sed_radio_oir}). 
Connecting those requires a break in the spectrum in the far-IR,
which is inconsistent with the  simplest jet models where 
main parameters follow the power-law radial dependences with constant indices \citep{BK79,Konigl81}. 
The strongest arguments against the jet contribution to the OIR, however, come 
from the rather hard OIR spectra of the flare. 
First, the hook-like behaviour on the $I-H$ CMD at the beginning of the flare (Fig.~\ref{fig_oir_cmd}b),
 a broken power-law flare spectrum which hardens towards red on MJD~51688  (greed solid line in Fig.~\ref{fig_spe_all}) 
as well as the hard flare spectrum with $\alpha>1$ in the end of the intermediate state  (see Section~\ref{sect:vvi} and Fig.~\ref{fig_oir_cmd}c) 
are not consistent with the jet spectrum extending from radio to the OIR band. 
Secondly, later in the hard state, the flare spectrum has $\alpha\sim 1$. 
If it were optically thick jet emission, the radio flux would be about five orders of magnitude below what 
was observed just a few days before. We find this highly improbable.  
Furthermore, it is difficult to understand how a substantial decrease in the accretion rate  
could lead to formation of the outflow which is optically thick up to the OIR band. 
The weakness of the jet in the OIR band also implies that the jet cannot possibly contribute to the X-rays. 

OIR flares were observed in a number of BH X-ray binaries. 
It was noticed that  the beginning of the flare nearly coincides with the radio brightening \citep{Kalemci13}. 
We find that the position of \source\ at the X-ray hardness--flux diagram 
at the times of quenching/recovery of the OIR non-thermal component  
(see Fig.~\ref{fig_xrays_lc}c) closely resembles the `jet line'  \citep{BMM11}. 
This coincidence can be interpreted in two alternative ways: 
(i) both OIR and radio emission  originate from one component (jet), 
or (ii) both are causally connected, but originate in different places. 
As we showed above, the first interpretation is at odds with the spectral  
evolution of the OIR flare. 
In the second interpretation, the OIR flare may originate in the geometrically thick hot flow 
(see below), while the radio is produced in the jet. The coincidence can be interpreted 
as a simultaneous appearance of the hot flow and the jet. 
This is indeed expected as in modern MHD simulations  the jet seems to form only 
if there is collimation by a geometrically thick accretion flow. 
In the soft state, when the cold, geometrically thin disc extends all the way to the innermost stable orbit, 
the jet is weak. 
 
\subsection{Hot accretion flow}
\label{sec:hotacc}
 
It was usually assumed that the electrons in the hot flow follow thermal distribution and in that case 
synchrotron emission in luminous BH X-ray binaries cannot possibly be of importance \citep{WZ00}, because 
of strong self-absorption.  
However, even if the flow contains a small, energetically unimportant tail of non-thermal electrons, 
situation changes dramatically, as the synchrotron luminosity increases by orders of magnitude \citep{WZ01,VPV13}. 
This results in two effects: first, the seed photons for Comptonization in the hot flow 
now can be dominated by the non-thermal synchrotron instead of the cold truncated accretion disc 
\citep{MB09,PV09} and, secondly, the synchrotron radiation can dominate the OIR emission from the BH \citep{VPV13,PV14}. 
The range of wavelengths where the hot flow  emits  is determined by its size: the larger is the 
truncation radius of the cold accretion disc, the lower is the frequency where synchrotron radiation is still not absorbed. 
The self-absorption (turn-over) frequency is \citep{VPV13}
\begin{equation}\label{eq:turn_over_exact}
 \nu_{\rm t} \approx 
  3\times 10^{15}  (B/10^6\,{\rm G})^{\frac{p+2}{p+4}}  \left[ \tau (p-1) \right]^{\frac{2}{p+4}} \ \mbox{Hz}, 
\end{equation}
where $B$ is the magnetic field strength,  $\tau$ is the Thomson optical depth of non-thermal electrons and 
$p$ is the slope the electron distribution. 

For the radial dependences $B\propto R^{-\beta}$ and $\tau\propto R^{-\theta}$ the   self-absorption  frequency scales as 
 $\nu_{\rm t}\propto R^{-[\beta(p+2)+2\theta]/(p+4)}$ 
 and the total OIR spectrum from a  hot flow  is composed of contributions of synchrotron peaks (in partially opaque regime) 
 coming from different radii. 
 The hot flow spectrum constitutes a power-law with the spectral  index \citep{VPV13}
\begin{equation}\label{eq:OIR_slope}
 \alpha_{\rm OIR}    =    \frac {5 \theta + \beta (2p + 3) - 2p - 8}{\beta (p+2) + 2\theta}.
\end{equation}
At frequencies above $ \nu_{\rm t} $ for the smallest, most compact zone of the 
accretion disc, the synchrotron spectrum is optically thin with $\alpha=-(p-1)/2$.  
If the optical depth of thermal electrons is high enough, thermal Comptonization can dominate 
completely over the optically thin synchrotron, the latter therefore might be invisible in the total spectrum. 
At frequencies below the self-absorption  frequency for the largest, most transparent zone,
the spectrum is optically thick with $\alpha=5/2$.  

Below we follow the hybrid hot flow scenario as described in details in \citet{VPV13}. 
We discuss now how it can explain the broad-band spectrum and the features seen in the CMD. 
We also discuss what are the implications for the physical parameters in the vicinity of the BH.

\subsubsection{Broad-band spectrum}

The broad-band OIR-to-X-ray spectra 
in the soft and intermediate states are fully consistent with the  standard accretion disc with the 
addition of the emission from non-thermal/hybrid electrons in the hard X-rays and the 
irradiated disc in the OIR band (see Fig.~\ref{fig_sed_broad_nuFnu}). The X-ray non-thermal emission can presumably be 
associated with the corona above the accretion disc, as in these states the hot inner flow 
either is non-existent or very small. Because of high luminosity of the standard disc, 
electron cooling is dominated by  Compton scattering, hence the synchrotron component from
the corona is negligible. Indeed, there are no any signatures of this emission in the OIR band. 

\begin{figure}
\centerline{\epsfig{file=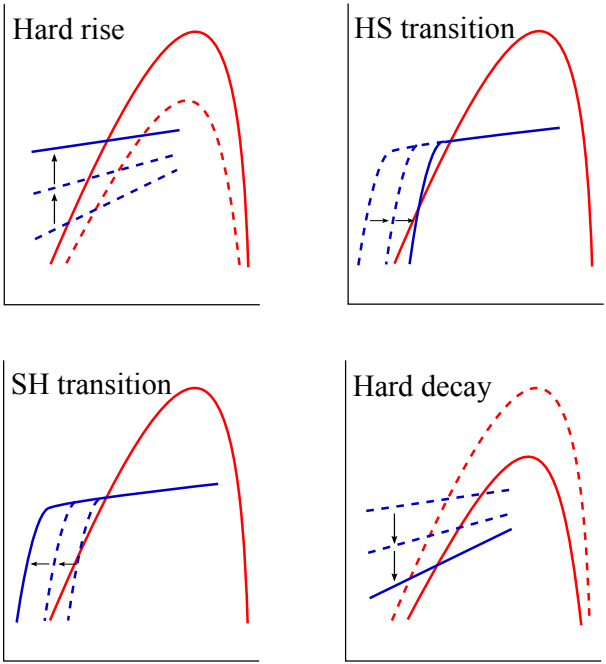, width=8cm}}
\caption{
The schematic evolution of the OIR spectra of the irradiated disc  and the flare components from the hot flow  during the outburst.  
At the rising phase of the outburst, in the hard state, the disc temperature is steadily growing together 
with the luminosity of the  non-thermal component. 
At the HS transition, the X-ray luminosity is nearly constant resulting in a nearly constant disc emission; 
the hot flow size shrinks affecting the synchrotron self-absorption frequency, which is increasing. 
At the soft state (not shown here), the non-thermal synchrotron component may not exist at all, because 
the hot flow has disappeared. 
At the SH transition, the hot flow appears and its size steadily grows resulting in a rapid evolution 
of the self-absorption frequency. In the following hard state, the X-ray as well as the disc luminosity drop; 
the non-thermal component decays even faster and becomes harder. 
}
\label{fig_sed_scheme}
\end{figure}

In the hard state, the whole  OIR to X-ray spectrum can be explained by a  
synchrotron self-Compton model where the synchrotron emission from non-thermal electrons 
is Comptonized by the thermal population of electrons \citep{PV09,VPV13}.    
At the beginning of the hard state (MJD~51688), the accretion rate  $\dot{M}$ is high, the synchrotron self-absorption frequency
of the innermost zone is at $\approx 3\times10^{15}$Hz (corresponding to $B\sim 10^6$~G 
and $\tau\sim1$, see equation (\ref{eq:turn_over_exact})),
while  $\nu_{\rm t}\sim3\times10^{14}$~Hz for the outermost zone of the hot flow 
resulting in a break to harder spectrum below that frequency. 
This explains a fact that the extrapolation of the X-ray power-law overpredicts the OIR flux (solid blue lines in Fig.~\ref{fig_sed_broad_nuFnu}). 
Later in the hard state, during the peak of the OIR flare (on MJD~51697), $\dot{M}$  and therefore $B$ and $\tau$  
have dropped, the hot flow has grown in size and  
$\nu_{\rm t}$ for both inner- and outermost zones have decreased. 
This results in a flat power-law spectrum in the OIR band 
(see dashed blue lines in Fig.~\ref{fig_sed_broad_nuFnu} and the 
SH transition in Fig.~\ref{fig_sed_scheme} for a schematic presentation of the spectral evolution).
Because at lower $\dot{M}$ the total optical depth has likely decreased, the synchrotron peak became more pronounced. 
Later in the hard state (on MJD~51706; dotted blue lines in Fig.~\ref{fig_sed_broad_nuFnu}), when the optical depth in the hot flow drops further, 
the OIR/X-ray spectrum cannot be represented by a broken power law any longer. 
Instead, it seems that there are two bumps in the OIR and the X-rays (similar to the spectra of blazars). 
Such double-peaked spectra are generally expected in the hot flow scenario at luminosities below a few per cent of the Eddington,
at low optical depths  (see fig.~6 in \citealt{NMQ98} and fig.~5 in \citealt*{VVP11}).

\subsubsection{Horizontal tracks in CMD}
 
The OIR spectrum of the hot flow constitutes a power law, which extends down to the self-absorption frequency $\nu_{\rm t}$ 
of the outermost parts. 
During the rising phase of the outburst, the cold accretion disc shrinks, gradually replacing the hot flow at smaller and smaller radii.
Thus $\nu_{\rm t}$ shifts to higher frequencies (see Fig.~\ref{fig_sed_scheme}).
Once the $\nu_{\rm t}$ crosses the  $H$ filter, the emission at these wavelengths becomes self-absorbed and further, 
even rather small increase of the transition frequency results in rapid decrease of the $H$-flux.
At the same time, the $V$ filter experiences minor changes, until the $\nu_{\rm t}$ reaches it.
At the decay phase of the outburst, the cold disc recedes, thus the transition frequency moves towards longer wavelengths, 
and the rapid flux increase in $H$ filter occurs.
This behaviour naturally explains the horizontal tracks we see in the CMDs (Fig.~\ref{fig_oir_cmd}).

The  $V$--$I$ CMD (Fig.~\ref{fig_oir_cmd}c) also shows a 
nearly horizontal movement but to the left of the blackbody track 
in the end of the SH transition corresponding to the hardening of the spectrum. 
This is a natural consequence of the appearance of a hard hot flow component which 
compensates the decreasing disc flux in  the $V$ filter but does not contribute much to the $I$ filter. 

It is interesting to note that  the horizontal tracks appear at different luminosities. 
These levels correspond to a factor of 1.35 difference in the outer accretion disc temperature,
while the X-ray  HS  and SH  transitions occur at luminosities that differ by a factor of 3. 
This is consistent with a simple relation $T_{\rm eff}\propto L_{\rm X}^{1/4}$. 
Thus, the horizontal tracks in the OIR CMDs and the X-ray hysteresis loop 
are both manifestation of the same phenomenon. 
We finally note that the hot flow emission in the OIR band disappears/appears in the 
intermediate states at the rising  and decaying stages of the outburst around MJD~51659 and 51683, 
respectively (see Figs~\ref{fig_oir_cmd}b,c), 
at a similar X-ray hardness (see Fig.~\ref{fig_xrays_lc}c). 
This strongly argues in favour of a similar truncation radius of the cold disc at these moments
in spite of a factor of three difference in X-ray luminosity.

\subsubsection{$I-H$ hook}

At the HS transition, when the cold disc truncation radius decreases, 
hot flow model predicts a quenching of the emission 
at longer wavelengths before that at shorter wavelengths.
For instance, quenching the emission in the $H$ filter results in a rapid colour change at an almost constant $I$-magnitude, 
as indeed seen in the $I$ versus $I-H$ CMD (Fig.~\ref{fig_oir_cmd}b).
Then, quenching of the emission in $I$ filter is expected, however, it is more difficult to see in this diagram, 
because both the magnitude and the colour change, at the same time the thermal emission is increased, partially compensating for the drop in $I$ filter.
The behaviour is tentatively seen in Fig.~\ref{fig_oir_cmd}(b), where the return to the blackbody track goes along 
the horizontal line (at constant $I$), then the $I$ flux also decreases.
However, due to large error bars and small number of data points, 
the evolution is also consistent with a decrease of the emission in both filters, $I$ and $H$.

The situation is different at the reverse,  SH transition.
With receding cold disc, the hot flow emission is recovered at shorter wavelengths before longer wavelengths.
We thus expect the evolution to proceed as follows: first, the spectrum follows a cooling blackbody, 
then the $I$ flux increases, thus both colour and magnitude change in the opposite direction (towards upper left in Fig.~\ref{fig_oir_cmd}b), 
then the $H$ flux increases and the horizontal track appears.
Such a behaviour is indeed seen as a hook at the $I-H$ CMD. 
The rise of the emission in the $I$ filter before the $H$ filter cannot be accounted for by the jet model.
It is thus a good proxy for discriminating between the jet and the hot flow scenario.

\subsubsection{Hot flow size and radial structure}

When the OIR flare becomes visible in the very beginning of the hard state at the decaying stage of the outburst, 
the non-thermal component has  spectral index $\alpha_{IH}\approx 1$, 
which is smaller than the value 2.5 corresponding to the self-absorbed spectrum. 
This indicates that $\nu_{\rm t}$ lies between the $I$ and $H$ filters.
Approximating the hot flow spectrum at these wavelengths with a broken power-law (see Fig.~\ref{fig_spe_all}), 
we obtain $\nu_{\rm t}\approx 3\times10^{14}$~Hz, 
from which we can infer the hot flow size at this moment of time.
Using equation~13 of \citet{VPV13}, which was derived under  an assumption of power-law radial dependences 
of the magnetic field strength and the electron density, we get 
\begin{equation}
 R = 3\times 10^{22}/\nu_{\rm t} \approx 10^8\ \mbox{cm} \approx 35 R_{\rm S},
\end{equation}
where $R_{\rm S}=2GM/c^2=2.7\times10^6$~cm is the Schwarzschild radius of the $9.1\msun$ BH.
This value should be considered as a rough estimate, to be improved by detailed spectral modelling.

We find that the non-thermal spectrum does not stay power-law of one particular spectral index, but is hardening towards the end of the flare.
According to equation (\ref{eq:OIR_slope}), $\alpha$ depends on indices $\theta$, $\beta$ and $p$.  
Let us assume that changes in $\alpha$ are predominantly caused by changes in the radial distribution of the optical depth, 
while $\beta$ is fixed at a value of $5/4$ for advective flows \citep{SK05}. 
For softest $\alpha=0.0$ and hardest $\alpha=1.0$, we obtain a change of $\theta$ from $\sim0.55$ to $3.0$ 
if we fix  $p=3$ (the precise value has minor effect on the results).
The value of  $\theta$ for the soft spectrum is similar to that found in ADAF flows \citep{KFM08}, 
however, the value obtained from the hard spectrum  is quite an extreme, as it requires a highly inhomogeneous configuration. 
We should remember, however, that here $\theta$ describes the distribution of non-thermal electrons, not 
their bulk. Thus, high values of $\theta$ might imply that electron acceleration is not efficient at large radii. 
If instead we choose to fix $\theta=0.5$, we get $\beta=1.3$ for the soft spectrum 
(which is close to the standard value in ADAFs),   
while for the hard spectrum we get $\beta=3.1$, again suggesting an extreme stratification of the magnetic field strength.
In reality, the structure of both the magnetic field and the optical depth is likely to change when 
the accretion rate decreases,  
with the general trend, apparent from the above exercise, such that the flow becomes more stratified with 
strong gradients.

\section{Summary}
\label{sect:summary}

We have carefully analysed the OIR and  X-ray  light curves of \source. From the e-folding times in different wavebands, 
we obtained  the outer accretion disc temperature in excess of 15\,000~K during the soft/intermediate states. 
Using this temperature together with the extreme OIR colours observed from \source, we put strong constraints on 
the extinction towards the source, $5.0\lesssim A_V\lesssim 5.6$. 
This allowed us to relate the observed OIR colours to the  slopes of the intrinsic spectrum.  

During the soft state, the OIR spectrum is well  described by the blackbody (associated with the 
irradiated disc) of 
exponentially decaying temperature and a constant normalization, while 
during the X-ray state transitions and the hard states, strong non-thermal flares are observed.  
By interpolating the decaying flux of thermal emission to the hard state, we have 
accurately separated the contribution of the flare from the disc. 
Importantly, the spectrum of the subtracted thermal component was shown to be 
consistent with the blackbody of the same normalization as before the flare. 
We further demonstrated how the spectrum of the  non-thermal component evolves. 
On the decaying phase of the outburst, the flare starts with an apparent break 
in the spectrum that is harder in the near-IR with $\alpha\approx1$. 
The spectrum then becomes a power law and softens to $\alpha\approx 0$ at the peak of the flare.
It  hardens again as the source fades. 
In the hybrid hot flow scenario, this hardening possibly 
indicates that the electron acceleration is not efficient at large radii at low accretion rates. 

The evolution of the spectral shape is consistent with the  hot accretion flow scenario,
where the size of the flow varies with the accretion rate.  
In the hard state, the flow is large and the OIR spectrum constitutes a power law. 
During the HS  transition, the flow collapses and the low-energy spectral cutoff 
(corresponding to the synchrotron self-absorption frequency) moves from the infrared to the UV. 
In the soft state, the hot flow may not exist at all and its non-thermal emission is quenched. 
At the reverse transition, the cold disc retreats and the hot flow grows in size leading to 
an increase of the contribution of the non-thermal  component from the hot flow 
to the OIR spectrum and strong reddening of the source.
This transition occurs at a lower accretion rate which manifests itself as the hysteresis loop in the X-ray hardness--flux diagram 
and as different  levels of the horizontal tracks in the OIR CMDs. 
We find that quenching and recovery of the hot flow OIR emission occurs at about the same X-ray hardness during the intermediate state. 
The position of the source at the X-ray hardness--flux diagram when this happens (the  hot flow line)  
seems to coincide with the jet line that marks the presence of the compact radio jet.  
This implies a close connection between the presence of the hot flow and the compact jet. 
We roughly estimate the hot flow size, at the moment when OIR flare starts to be visible in the light curves, 
to be $10^8\mbox{cm}\approx 35R_{\rm S}$.

\section*{Acknowledgements}

The work was supported by the Finnish Doctoral Programme in Astronomy and Space Physics (AV), 
the Academy of Finland grant 268740 (JP) and the Russian Science Foundation grant RNF 14-12-01287 (MGR).
We thank  David Russell for useful discussions and the OIR data on \source, 
Charles Bailyn for information about the filters used at the Yale telescope, and
Vitaly Neustroev  for information on photometric standards. 
This research has made use of data
provided by the High Energy Astrophysics Science Archive Research Center (HEASARC), which is
a service of the Astrophysics Science Division at NASA/GSFC and the High Energy Astrophysics
Division of the Smithsonian Astrophysical Observatory.


\label{lastpage}

\end{document}